\documentclass{eptcs}
\usepackage{latexsym,amssymb}
\usepackage{amsmath}
\newtheorem{defn}{Definition}
\newtheorem{thm}[defn]{Theorem}

\newtheorem{lem}[defn]{Lemma}
\newtheorem{cor}[defn]{Corollary}
\newcommand{\proofbeg}{\textbf{Proof. }}
\newcommand{\proofend}{\hfill $\dashv$}
\newcommand{\GA}{\mathfrak{A}}

\newcommand{\RA}{\mathcal{R}}
\newcommand{\SA}{\mathcal{S}}

\newcommand{\CS}{\otimes}

\newcommand{\TA}{\mathfrak{T}}

\newcommand{\FIN}{\mathsf{FIN}}
\newcommand{\SyS}{\mathsf{SyS}}
\newcommand{\SUB}{\mathit{SUB}}
\newcommand{\pr}{\mathcal{I}}
\newcommand{\gcp}{\mathit{gcp}}
\newcommand{\pref}{\mathit{pref}}

\title{Synchronous Subsequentiality and 
Approximations to Undecidable Problems}
\author{Christian Wurm
\email{cwurm@phil.hhu.de}
\institute{Universit\"at D\"usseldorf}}
\def\titlerunning{Synchronous Subsequentiality and Approximations}
\def\authorrunning{Christian Wurm}

\begin{document}
\maketitle
\begin{abstract}
We introduce the class of synchronous subsequential relations,
a subclass of the synchronous relations which embodies some 
properties of subsequential relations. If we take relations of 
this class as forming the possible transitions of an infinite
automaton, then most decision problems (apart from membership) still remain undecidable
(as they are for synchronous and subsequential rational 
relations), but on the positive side, they can be
approximated in a meaningful way we make precise in this paper. 
This might make the class useful for some applications, and 
might serve to establish an intermediate position in the trade-off
between issues of expressivity and (un)decidability.
\end{abstract}

\section{Introduction}

Automata play an important role for a huge number of tasks,
ranging from formal language theory to program semantics and
model checking for logical languages. The most widespread applications
have been found for finite automata; their properties and advantages are
well-known to an extent which makes comments unnecessary. There can be
several reasons to move from finite automata to infinite automata:
in formal language theory, it is expressive power, same holds for automata
as program semantics (a program might have at least in theory 
infinitely many possible states).
There might also be another reason:
for model checking, if we can compute a finite 
automaton directly from some formal (logical) language specifying 
the properties of the model (see for example \cite{courcelle:automata}),
1. the resulting automaton might be too
big to be effectively stored, or 
even if we can avoid this, it might happen that
2. though the resulting minimal automaton is manageable in size, its
construction involves intermediate steps with automata too large
to handle. For overview and motivation in logic, see
\cite{courcelle:automata}; in program semantics, this problem
is referred to as \textit{state explosion problem}, see \cite{modelchecking}.

So there are a number of reasons to use infinite automata. Their obvious
disadvantage is that we cannot simply write them down, as they are
infinite objects. What we rather do is the following: we specify a
recursive procedure by which an automaton constructs its own state 
set \textit{on the fly}, meaning it ``constructs" a certain state
only when reading a certain input. This leads to the theory of
infinite automata (see \cite{thomas:infinite} for an overview,
\cite{caucal:decidable},\cite{caucal:turing} for some important
results), and it is
this line of research on which we will investigate
here. To specify a class of infinite automata, the main work lies in
specifying a class of possible primitive transitions relations,
which are associated with the input letters, and, maybe of less
importance, classes possible sets of initial and final states. 

For classes of infinite automata, there is usually the following situation
of trade-off: Choice 1: we take a class of relations which is very restricted,
such as typically relations corresponding to transitions of pushdown automata
(henceforth: PDA) or slight extensions thereof (see for example
\cite{caucal:prefix}).
In this case, problems such as whether one state can be reached from another
one, and whether the language an automaton recognizes is non-empty are
decidable. The problem is that many computations cannot be simulated by
these transitions, and recognizing power remains quite limited.
Choice 2 is that we take a rather expressive class of relations such as
the rational (or regular) relations, computed by (synchronous) finite-state
transducers. These will provide sufficient expressive power for most purposes.
But on the downside, most problems -- reachability of states, emptiness of recognized
language etc. -- are undecidable. In between these two choices, there is
usually thought to be little to gain (see also \cite{thomas:infinite}, 
\cite{pierce:iterated}).
In particular, for classes of relations which do not have 
PDA-style restrictions, undecidability strikes very quickly,
so there is little use in investigating expressive subclasses of 
the rational relations. 

Yet this is exactly what we will do here: we will
investigate the class of synchronous subsequential relations, 
a rather large subclass of the regular relations, which embodies some 
properties of subsequential rational relations. It comes
as no surprise that also for this class, the reachability problem is
undecidable. Yet this class allows for an interesting way of 
approximating the reachability problem. This in turn allows to 
approximate other decision problems, which might make this
class interesting for some applications. The notion of
approximation we use is of some interest in itself: its intuitive
meaning is that we get arbitrarily close to a solution
of our binary problem, but we might
never reach it. In a binary problem, this is of course less
satisfying than in a numerical one. However, this can be put to use in various
ways, which correspond to various interpretations we can give to 
reachability problems.

\section{The General Setup: Self-constructing Automata}

\subsection{The Algebraic Setting: Self-constructing Automata}

One can present automata in many different ways, the most
standard one being probably the following:
an automaton as \textbf{state-transition system} (STS) 
is a tuple $\GA=(\Sigma,Q,\delta,F,I)$,
where $\Sigma$ is a finite input alphabet, $Q$ a set of states,
$\delta\subseteq Q\times\Sigma\times Q$ a transition relation,
$F\subseteq Q$ a set of accepting states, $I\subseteq Q$ the set
of initial states. 
As our main interest will be
in \textit{infinite} automata, we cannot simply write
down $Q$; same for $\delta$ and maybe $I,F$. The solution to this
is simple: we define $Q\subseteq\Omega^*$ for some finite 
alphabet $\Omega$ inductively by 1. $I\subseteq Q$, and 2. if
$q\in Q$, $a\in\Sigma$, then $\delta(q,a)\subseteq Q$,\footnote{Here
and throughout this paper, we treat relations as functions into
a powerset whenever it is convenient.} and
3. nothing else is in $Q$. 
Next, we simply
define $\delta$ as some computable relation,
$I,F$ as recursive sets, and we have a finite specification of an
infinite state machine. 


As in order to define an infinite STS, the main
work lies in specifying the transition relation,
we now introduce a more genuinely relational perspective, 
which amounts to a sort
of non-standard definition of automata, which we dub 
\textbf{self-constructing automata}, for short SCA.
This is more a notional
innovation than a substantial one, and we just adopt it
for convenience. 

Self constructing automata can be roughly conceived of as mappings
from the free  monoid $\langle \Sigma^*,\cdot,\epsilon \rangle$ to
the monoid $\langle \wp(\Omega^*\times\Omega^*),\circ,id_{\Omega^*}\rangle$,
where $\Omega^*$ is the free monoid over $\Omega$; 
$\circ$ denotes relation
composition; $id_{\Omega^*}$ denotes the identity relation on $\Omega^{*}$. 
We call the former the \textit{outer} algebra, the latter the \textit{inner} one. 
We will as much as possible stick to the convention of using
$\Sigma$ if something is supposed to belong to the outer algebra,
and $\Omega$ otherwise, though both designate finite alphabets throughout.
We define a semi-SCA as
a tuple $\langle \Sigma,\phi\rangle$, where $\phi$
is a map $\phi:\Sigma\rightarrow\wp(\Omega^*\times\Omega^{*})$, thus mapping
letters in $\Sigma$ onto relations over $\Omega$. It
is extended to strings in the usual fashion, where 
$\phi(aw)=\phi(a)\circ \phi(w)$; so $\phi$ is a homomorphism
from the outer algebra into the inner algebra. 
A word  $w\in\Sigma^*$ from the outer alphabet 
then induces a relation $\phi(w)\subseteq\Omega^*\times\Omega^*$.
We call the relations of the form $\phi(a):a\in\Sigma$
the \textbf{primitive  transition relations}.
To get a full automaton, 
we still need an \textit{accepting relation} of initial and accepting states.
One usually specifies a single initial and a set of accepting
states, yielding an accepting relation $\{x_0\}\times F$. 
As for us, acceptance will only play a minor role, we will
take a slightly more general convention and assume that
SCA specify an \textbf{accepting relation} 
$F_R\subseteq\Omega^*\times\Omega^*$, which might be a Cartesian product,
but need not be.
We denote the accepting relation by $F_R$, where the subscript is 
just a reminder that
we have a relation rather than a set of accepting states. 
Thus a full SCA is a tuple $\langle \Sigma,\phi,F_R\rangle$.
We can now define the language
recognized by an SCA. Let 
$\GA=\langle \Sigma,\phi,F_R\rangle$ be an
SCA; then $L(\GA):=\{w\in\Sigma^*:\phi(w)\cap F_R\neq \emptyset\}$.

For convenience, we always put $\phi(\epsilon)=id_{\Omega^*}$
(though this is not necessary in all cases).
One might at this point wonder how we deal with $\epsilon$-transitions.
There is an easy solution to that, by changing $\phi$.
Before we state it, we give the following definitions:
Given a set of relations $\{R_i:i\in I\}$, $I$ an arbitrary index set,
by $\{R_i:i\in I\}^\CS$ we define the smallest set such that:
\begin{enumerate}
\item $\{R_i:i\in I\}\subseteq\{R_i:i\in I\}^\CS$;
\item if $R,R'\in \{R_i:i\in I\}^\CS$, then
$R\circ R'\in \{R_i:i\in I\}^\CS$.
\item $id_{\Omega^*}\in \{R_i:i\in I\}^\CS$.
\end{enumerate}
In words, $\{R_i:i\in I\}^\CS$ contains $\{R_i:i\in I\}$, the
identity, and is closed under composition.
$[-]^\CS$ is obviously related to the Kleene-star closure, but for
composition of relations rather than for concatenation of strings. 
Be also careful to keep in mind that
we do not take the union over this closure, so 
$\{R_i:i\in I\}^\CS$ is \textit{not} a relation but  a
set of relations. We will sometimes refer to a set of the form
$\{R_i:i\in I\}^\CS$ as a \textbf{relation monoid}, as it is 
easy to see that this set is a monoid with operation $\circ$.
We will also need the union of this set and so  
define $\{R_i:i\in I\}^\oplus:=
\bigcup(\{R_i:i\in I\}^\CS)$. This is not exactly the smallest
reflexive, transitive relation containing every $R_i:i\in I$, because
it contains the full identity on $\Omega^*$; apart from this however it equals
the reflexive transitive closure. 
We urge the reader to be careful
to not confuse $\{R_i:i\in I\}^\CS$, a set of relations, with
$\{R_i:i\in I\}^\oplus$, a relation. Now assume we want a relation
$R_\epsilon$ to correspond to $\epsilon$, whereas for each letter
$a\in\Sigma$, we want a corresponding relation $R_a$. We can
simulate this in the algebraic setting by putting
$\phi(a):=(R_\epsilon)^\oplus\circ R_a\circ (R_\epsilon)^\oplus$.

Given some classes of relations $\RA,\RA_1,\RA_2$, 
by $\SA_\RA$ we denote the
class of all semi-SCA $\langle \Sigma,\phi\rangle$, where
for all $\sigma\in\Sigma$, $\phi(\sigma)\in\RA$. 
By $\SA_{\RA_1,\RA_2}$ we denote the class of all SCA
$\langle \Sigma,\phi,F_R\rangle$ where 
$\phi(\sigma)\in\RA_1$ and $F_R\in \RA_2$. A \textbf{class of SCA}
is a class of the form $\SA_{\RA_1,\RA_2}$ for some
classes of relations $\RA_1,\RA_2$.

\subsection{Decision Problems in the Relational Setting}

The most important decision problems in this paper can be states as follows:

\noindent
\textbf{Reachability problem}:
Given a class of automata $\SA$, 
is there an algorithm, which for any
$\langle\Sigma,\phi,F_R\rangle\in\SA$, $(x,y)\in\Omega^*\times\Omega^*$,
determines in a finite number of steps whether there is
$w\in\Sigma^*$ such that $(x,y)\in\phi(w)$?

\noindent
\textbf{Emptiness problem}: Given a 
class of automata $\SA$, is there an algorithm, 
which for any given automaton $\GA\in \SA$
determines in a finite number of steps whether $L(\GA)=\emptyset$?

\noindent
\textbf{Inclusion problem}: Given a 
class of automata $\SA$, is there an algorithm, 
which for any given automata $\GA,\GA'\in \SA$
determines in a finite number of steps whether $L(\GA)\subseteq L(\GA')$?

If there is such an algorithm, we say the 
problem is \textit{decidable} for $\SA$,
if there is no such algorithm, we say it is \textit{undecidable}. 
Note that in most cases (if $\SA$ contains the class of finite automata),
the inclusion problem subsumes the emptiness and universality
problem, as they amount to decide whether $L(\GA)\subseteq\emptyset$,
$\Sigma^{*}\subseteq L(\GA)$.
Obviously, there is a close connection between emptiness, reachability and
recursiveness of relation monoids; it is made precise
by the following lemma:

\begin{lem}\label{reach}
Let $\FIN$ be the class of finite relations.
Given a class of relations $\RA$, the following are equivalent:
\begin{enumerate}
\item The emptiness problem for $\SA_{\RA, \FIN}$ is decidable.
\item The reachability problem for $\SA_{\RA, \FIN}$ is decidable.
\item Every relation of the form
$\{R_i:i\in I\}^\oplus$, where $R_i\in\RA:i\in I$, $|I|<\omega$, 
is recursive.
\end{enumerate}
\end{lem}

\proofbeg
$1\Rightarrow 2$: Take a semi-automaton
$\langle\Sigma,\phi\rangle$, and ask whether there is
$w\in\Sigma^*$ such that $(x,y)\in\phi(w)$. This is the case
if and only if $L(\langle \Sigma,\phi,\{(x,y)\}\rangle)\neq\emptyset$,
which by assumption is decidable.

$2\Rightarrow 3$: Assume the reachability problem for
$\SA_{\RA, \FIN}$ is decidable, and for
$R_i\in\RA:i\in I$, $|I|<\omega$, 
ask whether $(x,y)\in (\{R_i:i\in I\})^\oplus$. We just
put $\Sigma:=I$, $\phi(i)=R_i$. Then we know whether
there is $w\in I^*$ such that $(x,y)\in\phi(w)$, which holds if
and only if $(x,y)\in (\{R_i:i\in I\})^\oplus$.

$3\Rightarrow 1$: Ask whether for $\langle\Sigma,\phi,F_R\rangle\in\SA_{\RA, \FIN}$,
$L(\langle\Sigma,\phi,F_R\rangle)=\emptyset$.
We take  $(\{\phi(a):a\in \Sigma\})^\oplus$, which is recursive
by assumption. Now we have
$L(\langle\Sigma,\phi,F_R\rangle)=\emptyset$ if and only
if for all $(x,y)\in F_R$, 
$(x,y)\notin(\{\phi(a):a\in \Sigma\})^\oplus$. As $F_R$ is
finite, this can be effectively checked for all members.
\proofend

Of course it is a rather significant restriction to only
consider finite accepting relations. But keep in mind
that for many (most?) automata, we can boil it down
to this case without loss of recognizing power, provided we allow 
$\epsilon$-transitions.\footnote{We need not talk of
FSA, but this holds also for PDA,TM and many intermediate classes.}

\section{Synchronous Subsequential Relations}

\subsection{Synchronicity and Subsequentiality}

The intuitive notion of synchronicity is that we use finite-state 
transducers which do not have $\epsilon$-transitions. A (finite-state) transducer is
a tuple $\langle Q,\Sigma,F,q_0,\delta\rangle$, where $F\subseteq Q$,
$q_0\in Q$, $Q,\Sigma$ are finite, and 
$\delta\subseteq Q\times\Sigma\cup\{\epsilon\}\times\Sigma\cup\{\epsilon\}\times Q$. 
Transducers are
based on the operation $\cdot$, where $(a,b)\cdot(c,d)=(ac,bd)$. We extend
$\delta$ to $\hat{\delta}$ 
by $\hat{\delta}(q,a,b)=\delta(q,a,b)$, and
$\hat{\delta}(q,aw,bv)=\{\hat{\delta}(q',w,v):q'\in\delta(q,a,b)\}$;
$L(\GA)=\{(w,v):\delta(q_0,w,v)\cap F\neq\emptyset\}$. If
$R$ is recognized by a finite state transducer having transitions
of the form $(\epsilon,a)$ and/or $(a,\epsilon)$, then it is \textbf{rational}.
The main advantage of synchronicity is that without $\epsilon$
appearing in components, the operation $\cdot$ still gives
rise to a free monoid: each term has a unique maximal decomposition. 
This property gets lost with $\epsilon$: $(a,\epsilon)\cdot(\epsilon,b)=
(\epsilon,b)\cdot (a,\epsilon)$. So this means that synchronous
transducers can be reduced to simple string automata for most
properties, whereas transducers in general cannot.

However, under this strict definition disallowing $\epsilon$-transitions, 
the corresponding relations are rather useless
for infinite state transition systems, as from each state, only
a subset of the finite set of equally long states is reachable.
So one uses a more liberal definition: we allow $\epsilon$-transitions to occur,
but only if in the component in which they occur, there are no
more other letters to come. By $|w|$ we denote the length of a string,
by $w^k$ its $k$ iterations.
Put $\Omega_{\perp}:=\Omega\cup\{\perp\}$, 
for $\perp\notin\Omega$. The \textbf{convolution}
$\odot  (w,v)$ of a pair of strings 
$(w,v)\in \Omega^{*}\times\Omega^{*}$ is defined by
$\odot  (w,v):=(w\!\!\perp^{max\{0,|v|-|w|\}},v\!\!\perp^{max\{0,|w|-|v|\}})$. So
in simple words: we take the shorter word of the two
and add $\perp$-symbols to make
it as long as the other. 
The convolution of a relation
$R\subseteq (\Omega^{*})^{2}$ is defined as 
$\odot R:=\{\odot(w,v)$:
$(w,v) \in R\}$.
A relation $R\subseteq \Omega^{*}\times\Omega^{*}$ is \textbf{regular}, 
if there is an ($\epsilon$-free) finite state transducer
over $(\Omega_{\perp})^{2}$ recognizing $\odot R$;
we denote this class by $REG$. 
This is the notion of synchronicity we will use here.
Regular relations as defined here form a proper subclass of the
relations defined by finite state transducers in general, but the 
restriction comes with a huge gain:
$REG$ is closed under Boolean operations, whereas the rational 
relations are not closed under intersection and complement
(see \cite{berstel:transductions}). This makes the regular relations
the more natural choice for application for example in logic (see \cite{rubin:automata}).

It is easy to see
that $\SA_{REG}$ has an undecidable reachability problem, as Turing
machine transitions are regular. A highly non-trivial result
shows that $\SA_{REG,\FIN}$ recognizes exactly 
the context-sensitive languages (proved in
\cite{rispal:synchronized}).

The intuitive notion of \textbf{subsequentiality} is the following:
relations on strings are subsequential, if computations depend
on prefixes that have already been read, but \textit{not} on
the part of the input which has not been read yet 
(in the sequel, we refer to this as ``future").
It is easy to define this concept for rational relations:
let $\TA=\langle Q,\Sigma,\delta,q_0,F\rangle$ 
be a transducer (possibly using $\epsilon$-transitions)
which is \textit{total}, that is, for every
$a\in\Sigma,q\in Q$, there is some $(q,a,a',q')\in\delta$.
Then $\TA$ is \textbf{subsequential}, if $Q=F$. In this sense, 
there is no way to discard any 
computation we have made so far. A relation $R$ is \textit{subsequential}
if it is computed by some subsequential transducer. We denote this
class by $\SUB$. 
%
%
%
%
%
The reachability
problem for $\SA_{\SUB}$ is again \textit{undecidable}. Though this result
is not literally stated at any point we know,
it easily falls off from results in \cite{fernau:iterated}
(who prove that iterated subsequential transduction generates
languages which are not recursive). Moreover, as below we show
a stronger result (theorem \ref{undec}), 
we omit the formal statement and proof of this one.
What we now do is to pair the concept
of subsequentiality with the concept of synchronicity.

\subsection{Synchronous Subsequential Relations}

It is not reasonable to simply define a
synchronous subsequential relation as being computed by
a synchronous transducer with only accepting states: 
because in this case still
computations depend on ``the future": simply because
whether we can map (or read) some $\epsilon$-pendant $\perp$
depends on the symbols \textit{which are still to come} (as $\perp$
can only occur final); so in 
some cases, we would have to discard computations which have been
executed, which contradicts the essence of subsequentiality. 
So we have to use another definition of synchronous subsequential
relations, taking a detour over (one-sided)
infinite words. Let $\Omega$ be an alphabet; we denote
the set of infinite words over $\Omega$ by $\Omega^\omega$;
formally, we can think of them as functions $\mathbb{N}\rightarrow\Omega$;
so they have the form $a_1a_2...a_n...$: we have a first letter,
every letter has an immediate successor, and every letter is preceded
by finitely many letters. For clarity, we will designate infinite
words with $\overline{x},\overline{y}$ etc.
Furthermore, we need one special letter 
$\square$, which we assume to be in all our alphabets
$\Omega$ over which we form infinite words (please read this as a dummy
for an arbitrary letter, which however has to be explicitly designated). 
We can get back from infinite words to finite words via
the following map
$\eta:\Omega^\omega\rightarrow \Omega^\omega\cup\Omega^*$: 

$\eta(a_1a_2...a_n...)=\begin{cases}
a_1\eta(a_2...a_n..)$, if
$a_2...a_n...\notin\{\square\}^\omega$, and$\\
\epsilon$ otherwise.$\end{cases}
$

So for $\overline{x}\in\Omega^\omega$, 
$\eta(\overline{x})\in\Omega^*$ iff all but finitely many letters
are $\square$, and $\eta(\overline{x})=\overline{x}$ otherwise. 
We now complete our definition of synchronous subsequential relations
as follows: let $\TA:=\langle Q,\delta,\Sigma,q_0\rangle$ be a total (wrt. input $q,a$)
$\epsilon$-free
transducer without accepting states. We put
$L^\omega(\TA):=\{(a_1a_2...,b_1b_2...)\in\Omega^\omega:
\delta(q_0,a_1,b_1,q)\in\delta$, and for all $i\in\mathbb{N}$,
there is some $q\in \hat{\delta}(q_0,a_1...a_{i},b_1...b_i)$ and $q'$ such that
$(q,a_{i+1},b_{i+1},q')\in \delta\}$.
We say $\TA$ is \textbf{synchronous subsequential}, if for 
every $\overline{w}\in \Omega^\omega$,
if $\eta(\overline{w})\in \Omega^*$, 
$(\overline{w},\overline{v})\in L^\omega(\TA)$, then
$\eta(\overline{v})\in \Omega^*$. We thus want for
any input with a finite $\eta$-image the output to have a
finite $\eta$-image as well. Technically, this can be easily 
implemented by making sure that after some number of 
input of $\square$-symbols, we end up in states which only
give $\square$ as output given a $\square$-input. Call these states \textbf{final}
(not to be confused with accepting). 
The motivation for this definition
is the following: we can use infinite words with finite 
$\eta$-image as if they were finite words (in the sense
of effective computation). 

\begin{defn}
We say that a relation $R\subseteq\Omega^\omega\times\Omega^\omega$ is in $\SyS$, the class of synchronous
subsequential relations, if there is a synchronous subsequential 
transducer $\TA$ such that $R=\{(\overline{w},\overline{v}):
(\overline{w},\overline{v})\in L_\omega(\TA)\ \&\ \eta(\overline{w})\in\Omega^*\}$.
\end{defn}

The trick is the following: as we do not have accepting states, there is no
additional complication when considering infinite words as opposed to finite
words, as complications arise from mode of acceptance. As we do not have any
$\epsilon$-transition, if we restrict ourselves to have a finite input (modulo $\eta$), that is
sufficient to ensure we have a finite output (mod  $\eta$). Note that 
$\square$ plays a double role in this definition: if it is followed by some other symbol
$a\neq\square$, then it is a symbol just like any other. If it is followed by
$\square$-symbols only, it basically plays the role of $\perp$ in the regular relations,
that is, it is a dummy-symbol mapped to $\epsilon$. Note that this ``trick"
is necessary to pair synchronicity with subsequentiality, as we do need a 
dummy symbol which gets deleted if final, and at the same time we need to
ensure that transitions are total.
An easy way to understand the relation of finite/infinite words is by analogy
with the real numbers: a number as $3.05$ is in some sense a ``shorthand" for the
number $3.050000...$; we can cut away final $0$s, but only if there is no other number
to come. 

For purposes of practical computation, this oscillation between finite and infinite
words does not pose any problem: we can just take a finite input word $w$
and take all the outputs we reach with some input
$w\square^{n}$ with which go to a final state. The resulting set of outputs
(mod $\eta$) is exactly the output for the infinite input word $w\square^{\omega}$.
So if we have a specific finite input word, we can compute the output by means of
a finite transducer with some accepting states (corresponding to the final states). However, this
does \textit{not} hold in the more general case where we have an infinite set of input words.
This is because when reading an input 
$w\square^{n}$, which is a prefix of an infinite word $\overline{w}$, we can 
\textit{never be sure} whether $\eta(\overline{w})=w$, so we can never actually
stop the computation.
This happens for example in relation composition, and one 
might consider this problematic. However, in this case 
we cannot write down the set of outputs anyway; and so we 
have to compute a finite representation of the infinite output set (or relation). 
The most important lemma to ensure applicability of $\SyS$ for infinite automata
is the following:

\begin{lem}
If $R,R'\in\SyS$, then $R\circ R'\in\SyS$.
\end{lem}

\proofbeg
We can easily show this by the standard transducer construction: given
synchronous rational transducers $\GA_1=(Q_1,\Sigma_1,q_0^1,\delta_1),
\GA_2=(Q_2,\Sigma_2,q_0^2,\delta_2)$,  we simply
construct $\GA_3=(Q_1\times Q_2,(q_0^{1},q_0^{2}),\Sigma_1\cup\Sigma_2,\delta_3)$,
where $\delta_3$ is defined as follows:
$((q,r),a,b,(q'r'))\in\delta_3$, iff $(q,a,c,q')\in\delta_1$ and
$(r,c,b,r')\in\delta_2$. It can be easily checked (by induction on word length\footnote{One
might correctly observe that an induction is insufficient for the case of infinite words.
However, we can reduce this case to the finite case as sketched above. We skip details
as the construction is standard.})
that this construction works.
\proofend

\begin{lem}
$\SyS\subsetneq REG$.
\end{lem} 

\proofbeg
1. $\subseteq$. Assume $R\in\SyS$. Take the synchronous
subsequential transducer $\TA$ recognizing $R$. Make an additional, disjoint copy
of the final states that is 1. accepting, 2. absorbing (no leaving arcs ) 
and 3. where you change $\square$ to $\perp$. This is synchronous and does the job.

2. $\neq$. Every finite relation is synchronous. A finite relation which is not synchronous
subsequential is $\{(aac,aa),(aad,ab)\}$. This is because $\SyS$-transducers have to be
total, and if we compute a prefix $(aa,aa)$, we have an input letter $d$, there must be
some output word corresponding to input $aad$.
\proofend

By $\pref(w)$ we denote all prefixes of $w$.
Relations in $\SUB$ generally have the \textit{prefix property}: 
if $(w,v)\in R$, then for every $w'\in \pref(w)$, there is 
$v'\in \pref(v)$ such that $(w',v')\in R$ (proof is straightforward).
$\SyS$ does not have this property, as we see below:

\begin{lem}
$\SyS\not\subseteq \SUB$
\end{lem}

\proofbeg
Take simply a relation $R$ which computes the
identity on all words in $\{\square,a\}^*$, which end with $a$. 
Obviously, $R\in \SyS$; to see that $R\notin \SUB$, just note that
$R$ does not have the prefix property.
\proofend

The underlying reason is given by the sketch
above: up to some extent, the computations of a synchronous subsequential transducer
depend ``on the future", namely as regards the question whether 
$\square$ is mapped to $\epsilon$ or not. $\SUB\not\subseteq\SyS$ is
obvious, as $\SUB\not\subseteq REG$. Note moreover that in the case of
a one-letter alphabet, all $\SyS$-relations over this alphabet are trivial.

\subsection{$\SyS$ and Program Semantics: an Example}\label{program}

To exemplify the meaning of $\SyS$-relations in terms of computational power,
let us give the states of infinite automata the following interpretation:
A state, being a finite string, represents a current state of a program. This program
state in turn is a vector containing
values of all declared objects (this comprises all variables, memory, registers,
program counters etc.). We can thus think of \textit{regions} of the string
as values of variables. For example, we might have
$\overbrace{a_1a_2a_3}^{\text{value of }x_1}\overbrace{a_4}^{\text{boundary}}\overbrace{a_5...a_8}^{\text{value of }x_2}...$.
Now assume the program computes the following function for all $i$ such that
$x_{i+1}$ is a declared object: 

\begin{center}
$f(x_{i+1}):=\begin{cases}f_1(x_{i+1})$, if $x_1,...,x_i$ satisfy condition $C_1\\
f_2(x_{i+1})$ if $x_1,...,x_i$ satisfy condition $C_2\\
\hdots
\end{cases}$
\end{center}
where $f_1,f_2,..$ are more basic computations and $C_1,C_2,...$ are exhaustive. 
This kind of computations is exactly what 
relations in $\SyS$ do.
An intermediate sequence of $\square$ symbols
can be thought  of as saying: ``variables are not instantiated", 
just giving them some
default value. We can thus perform computation
steps where the computation on variable $x_{i+1}$ depends on the values of
(and computations performed on) earlier variables, but not on those
which have higher index. This is due to the restriction to subsequentiality. 
The restriction regarding synchronicity concerns a restricted ability to insert new
variables into the existing order (it is clear that the order of variables is of crucial
importance in this model): in one computation step, we can insert only a globally
bounded number of new variables into the existing hierarchy, but not
a arbitrary number (say, after any existing variable, insert a new one into the
hierarchy). Moreover, it poses some bounds to value changes of variables.

We can thus say: $\SA_{REG}$ corresponds to computations where values of all variables
are computed in dependence of one another, $PDA$ correspond to computations where
only \textit{one} variable can change value in the course of a computation step.
$\SA_\SyS$ then corresponds to the intermediate situation where there is 
a linear hierarchy (i.e. sequence)
of variables, where the computations of variables higher up in 
the hierarchy (i.e. having lower index) have
impact on computations on lower variables, but not vice-versa.
So there are some restrictions; still it is easy to see that a huge number of computations
can be performed in this way!

\section{Some Formal Properties of $\SyS$}

We now scrutinize some more properties of $\SyS$.
We denote by $\diamond$ the product of two infinite strings, which is
defined as follows: $(a_1a_2....)\diamond (b_1b_2...)=(a_1,b_1)(a_2,b_2)...$.
So if $\overline{w}\in(\Omega_1)^\omega$, $\overline{v}\in(\Omega_2)^\omega$,
then $\overline{w}\diamond\overline{v}\in(\Omega_1\times\Omega_2)^\omega$.
We lift this operations to sets in the canonical fashion and extend
it canonically to relations, such that 
$R\diamond R'=\{((\overline{x}\diamond \overline{x'}),
(\overline{y}\diamond \overline{y'})):(\overline{x},\overline{y})\in R,(\overline{x'}
,\overline{y'})\in R'\}$.
It is easy to see that we could also define this operation for
finite words, but it would require some additional definitions to
``synchronize" them.  There is one thing we have to take care of:
we have used a symbol $\square$ with a special meaning
in the definition of $\SyS$.
We define a map $f_\square$, which is a string homomorphism
defined by $f_\square(\square,\square)=\square$, and $f_\square(x)=x$
otherwise. Again, we lift this map to sets and
relations in the canonical fashion. 
A simple, yet important lemma is the following:

\begin{lem}
If $R,R'\in \SyS$, then $f_\square(R\diamond R')\in\SyS$.
\end{lem}

\proofbeg
Take $\TA_1=(Q,\Sigma_1,q_0,\delta_1),\TA_2=(R,\Sigma_2,r_0,\delta_2)$. 
We put $\TA_3=(Q\times R,f_\square(\Sigma_1\times\Sigma_2),(q_0,r_0),\delta_3)$,
where $\delta_3$ is defined by:
$((q,r),f_\square(a,b),f_\square(a',b'),(q',r'))\in\delta_3$ iff
$(q,a,a',q')\in\delta_1$,$(r,b,b',r')\in\delta_2$.
This is a standard construction for FSA, so we leave its
verification to the reader.\footnote{A very similar construction
is performed when constructing the automaton recognizing the
intersection of two regular languages.}
\proofend

\begin{lem}
Let $\mathcal{R}_1,\RA_2$ be classes of relations. 
If $\mathcal{R}_1,\RA_2$ are closed under $\diamond$,
then the class of languages $L(\mathcal{S}_{\RA_1,\RA_2})$ is 
closed under intersection.
\end{lem}

\proofbeg
We can show this by construction: take 
$\GA_1:=\langle\Sigma_1,\phi_1,F_{R1}\rangle$,
$\GA_2:=\langle\Sigma_2,\phi_2, F_{R2}\rangle$.
Without loss of generality, we 
assume that $\Sigma_1=\Sigma_2$.\footnote{We can always enlarge
alphabets with a new letter $\sigma$, putting $\phi(\sigma)=\emptyset$.
Of course we have to assume $\emptyset$ is in any class of relations.}
We construct a new automaton
$\GA_3:=\langle\Sigma_1,
\langle\phi_1,\phi_2\rangle, (F_{R1}\diamond F_{R2})\rangle$,
where $\langle\phi_1,\phi_2\rangle:\Sigma_1\rightarrow 
(\Omega_1\times \Omega_2)^\omega\times (\Omega_1\times \Omega_2)^\omega$
is defined by $\langle\phi_1,\phi_2\rangle(w)=\phi_1(w)\diamond\phi_2(w)$.
It is easy to see that $L(\GA_3)=L(\GA_1)\cap L(\GA_2)$. 
\proofend

It is also easy to see that in particular the class $\FIN$ of 
finite relations is closed under
$\diamond$. Moreover, regarding $\SyS$, the map
$f_\square$ does not affect recognizing power, so it remains
a technical detail we will ignore in the sequel. Next we prove the following:

\begin{lem}\label{cfl}
Let $L$ be a CFL. Then there is a $\GA\in\SA_{\SyS,\FIN}$ such that 
$L=L(\GA)$.
\end{lem}

\proofbeg
We show how to encode a PDA in $\SyS$. The argument is 
conceptually simple yet 
tedious in full formality, so we give a 
rather informal explanation.

We encode stacks (and control states) as strings, and assume we read them from
bottom to top. Moves which consist in pushing something onto the stack
are unproblematic: we just have to make sure that
after reading $(x,x)$ we have a transition $(\square,y)$,
and this is sufficient if we make sure for all reachable states
that if we encounter $\square$
at some point, then only $\square $ is to follow in that component. 

The problem comes with pop-moves: reading $\overline{x}$ from left
to right, we cannot tell whether a certain
symbol is final in $\eta(\overline{x})$. We therefore proceed as follows:
a set of popping PDA-transitions $\{(xa,x):x\in\Omega^*,a\in\Omega\}$ is encoded
by a set of transitions $\{(xab_1b_2...,x\square c_1c_2...):x\in\Omega^*$
and for all $i\in\mathbb{N}$, if $b_i\neq\square$, then $c_i=\blacksquare$
and $c_i=\square$ otherwise$\}$
(where $\blacksquare$ does not figure in the original stack
alphabet). This relation is clearly in $\SyS$. Note that for all states
not containing $\blacksquare$, this preserves the property that if
we encounter $\square$
at some point, then only $\square $ is to follow in that component. 

Thereby, we use $\blacksquare$ as an ``absorbing symbol", that is, a symbol
which can never be eliminated from a string, and which does not figure in any
accepting state of the automaton $\GA\in\SA_{\SyS,\FIN}$. We use it to exclude all transitions 
which do not correspond to ``well-formed" stack moves 
from reaching an accepting state. 
This creates some additional non-determinism wrt. the PDA, but makes sure all
$\blacksquare$-free states have been reached by legitimate
PDA-moves.
\proofend

These lemmas, taken together, already have a strong immediate
consequence:

\begin{thm}\label{undec}
Given $R_i\in \SyS:i\in I$, $|I|<\omega$, it is undecidable 
whether $(x,y)\in \{R_i:i\in I\}^\oplus$.
Equivalently, the reachability problem for $\SA_{\SyS}$ is undecidable.
\end{thm}

\proofbeg
Assume conversely the problem is decidable. Then equivalently,
the problem whether $L(\GA)=\emptyset$ is decidable for all
$\GA\in\SA_{\SyS,\FIN}$. Now we take two
context-free languages $L_1,L_2$, accepted by two PDA by empty stack.
We encode the two PDA into $\GA_1,\GA_2\in\SA_{\SyS,\FIN}$. 
By lemma \ref{cfl}, we have $\GA_3\in\SA_{\SyS,\FIN}$, such that
$L(\GA_3)=L(\GA_1)\cap L(\GA_2)$. 
By assumption, we can decide whether $L(\GA_3)=\emptyset$, which holds
if and only if
$L_1\cap L_2=\emptyset$. But this is well-known to be an
undecidable problem for two context-free languages - contradiction.
\proofend

\begin{cor}
The emptiness, universality  and inclusion problems for $\SA_{\SyS,\FIN}$
are undecidable.
\end{cor}

So we do have a negative result for reachability in the first place.
The proof also tells us how powerful $\SyS$ is despite its restrictions, as we can recognize
any intersection of context-free languages, and so if we allow $\epsilon$-transitions,
we can already recognize every recursively enumerable language (it is also well-known
that by simulating two pushdowns we can easily simulate a Turing machine).
However, we can show that there is a way to approximate its decision
problem in a way to be made precise.

\section{Decompositions and Approximations}

We define the maps $\tau_n:n\in\mathbb{N}$ by
$\tau_n(a_1a_2...,b_1b_2...)=(a_n,b_n)$. 
For $R$ a relation, we put $\tau_n(R)=\{\tau_n(\overline{w},\overline{v}):
(\overline{w},\overline{v})\in R\}$ (again, for simplicity we give
the definition only for infinite words). Note that 
$\tau_n$ is not necessarily a homomorphism for 
relation composition: Put $R_1=\{(ab...,ba...)\},R_2=\{(bb...,aa...)\}$;
then $\tau_1(R_1)\circ \tau_1(R_2)=\{(a,a)\}$, whereas
$\tau_1(R_1\circ R_2)=\emptyset$. 
However, we have one inclusion: for all $n\in\mathbb{N}$,
 $R,R'\in\{R_i:i\in I\}^\otimes$,
$\tau_n(R\circ R')\subseteq\tau_n(R)\circ \tau_n(R')$, as is easy
to see. 
%
%

These notions are related
to the notion of direct and subdirect decompositions, fundamental to
universal algebra (see \cite{burris}).
We can say a set $\{R_i:i\in I\}$ is \textbf{directly decomposable},
if for all $n\in\mathbb{N}$, $\tau_n$ is a $\circ$-homomorphism
for $\{R_i:i\in I\}^\otimes$, and in addition,
%
$\prod_{n<\omega}\{\tau_n((R_i:i\in I)^\otimes\}=\{R_i:i\in I\}^\otimes$
(here, $\prod$ refers to the operation $\cdot$ -- the pointwise
concatenation of pairs -- which is extended to sets).
%
We can say $\{R_i:i\in I\}^\otimes$ is
\textbf{subdirectly decomposable}, if for all $n\in\mathbb{N}$,
$\tau_n$ is  a $\circ$-homomorphism. One can readily
check that this is sufficient for the usual conditions of 
(sub)direct products to hold. 
We also define the related family of maps $\sigma_n:n\in\mathbb{N}$
by $\sigma_n(\overline{x},\overline{y})=\prod_{i=1}^{n}\tau_i(\overline{x},\overline{y})$.
The main motivation for considering $\SyS$ is the following
result:

\begin{thm}\label{approx}
Assume $\langle \Sigma,\phi\rangle\in \SA_{\SyS}$. Then for all
$n\in\mathbb{N},a,b\in\Omega,x,y\in\Omega^{*}$, the sets
$\{w\in\Sigma^*:(a,b)\in\tau_n(\phi(w))\}$ and
$\{w\in\Sigma^*:(x,y)\in\sigma_n(\phi(w))\}$
are regular languages.
\end{thm}

We prove this only for $\sigma$, as the two proofs are almost identical.
If we have a function $f:M\rightarrow N$, then for $X\subseteq M$, we write
$f[X]:=\{f(x):x\in X\}$.

\proofbeg
Take an arbitrary fixed $n$. For $\overline{x},
\overline{y}\in\Omega^\omega$, we write
$\overline{x}\sim_n \overline{y}$, if $\overline{x}=z\overline{x'},
\overline{y}=z\overline{y'}$ with $|z|=n$. This is obviously
an equivalence relation. The notion of synchronous
subsequentiality now allows for the following argument:
If $\overline{x}\sim_n\overline{x'}$, then for any $w\in\Sigma^*$,
we have $(\overline{x},\overline{y})\in\phi(w)$, if and only if we have
$(\overline{x'},\overline{y'})\in \phi(w)$ for some $\overline{y'}:
\overline{y'}\sim_n \overline{y}$. 
Moreover, if $\overline{x}\sim_n\overline{x'},\overline{y}\sim_n \overline{y'}$, 
then $\sigma_n(\overline{x},\overline{y})=\sigma_n(\overline{x'},\overline{y'})$.

By this, we can obtain a congruence: 
Define $\phi_{\sim_n}$ by $\phi_{\sim_n}(w)=[\phi(w)]_{\sim_n}$,
that is, the set of $\sim_n$-equivalence classes of $\phi(w)$.
It is then clear that we have $\sigma_n\circ \phi(w)=
\sigma_n\circ\phi_{\sim_n}(w)$, if we choose an arbitrary
representative for each equivalence class.
Moreover, the monoid $\phi_{\sim_n}[\Sigma^*]$
is \textit{finite}, simply because $\Omega^\omega$ modulo $\sim_n$
has only finitely many equivalence classes. So $\phi_{\sim_n}$
is a map from $\Sigma^*$ into a finite set of relations,
and hence for every element, the pre-image consisting of the words
mapped to relations containing this element, is a regular
set (this is a fundamental result of the algebraic theory
of finite automata, see \cite{berstel:transductions}). 
Hence $\{w\in\Sigma^*:(x,y)\in\sigma_n(\phi(w))\}$ is regular.
\proofend

To see how $\sigma,\tau$ are connected, consider that 
we have 
\begin{center}
$\{w\in\Sigma^*:(a_1...a_n,b_1...b_n)\in\sigma_n(\phi(w))\}=
\bigcap_{i\leq n}\{w\in\Sigma^*:(a_i,b_i)\in\tau_i(\phi(w))\}$
\end{center}
As regular languages are closed under
intersection, we thus obtain a regular language 
for any finite sequence of indices. 
Moreover, the emptiness, universality and inclusion problem for regular languages
are decidable, and our computation effective. 
We now take the following convention:
given $x,y\in\Omega^*$,
we put $f_\omega(x,y)=
\{(x\overline{x},y\overline{y}): \overline{x},\overline{y}\in \Omega^\omega,\eta(\overline{x}),\eta(\overline{y})\in\Omega^{*}\}$.
Then the main result on approximation reads: 

\begin{cor}
For any $\langle\Sigma,\phi\rangle\in\SA_{\SyS}$, $(x,y)\in \Omega^*\times\Omega^*$, 
the set $\{w:\phi(w)\cap f_\omega(x,y)\neq\emptyset\}$
is a regular set which can be effectively computed. 
\end{cor}

This means that we can approach the problem
of emptiness/reachability by calculating the 
language for longer and longer prefixes of
the infinite pair $(\overline{x},\overline{y})$, 
which yields smaller and smaller regular
languages. $\SyS$ can be characterized in another very
natural fashion:

\begin{lem}\label{comm}
Let $\{R_i:i\in I\}$ be a set of relations, such that for all $i\in I$,
$R_i\in \SyS$. Then $\sigma_n(\{R_i:i\in I\}^\oplus))=(\sigma_n[\{R_i:i\in I\}])^\oplus$.
\end{lem}

\proofbeg
Follows directly from the conditions of 
synchronicity, subsequentiality and totality.
\proofend

This is in some sense the essence of synchronous subsequentiality.
The converse implication of lemma \ref{comm} 
is however wrong: just take any singleton set
$\{R\}$, where $R\circ R=\emptyset$. This satisfies the above equation,
and it is easy to find $R\notin \SyS$ for this.

\section{The Limits of Approximation}

One might think that this allows for \textit{effective} approximation
of the reachability problem in the sense that if $(\overline{x},\overline{y})
\notin \{\phi(a):a\in\Sigma\}^\oplus$, 
then there is some finite $(x,y)$ such that 
$(\overline{x},\overline{y})\in f_\omega(x,y)$ and
$\{\phi(a):a\in\Sigma\}^\oplus\cap f_\omega(x,y)=\emptyset$.
In this case, the approximation would always succeed in the case
that actually $(\overline{x},\overline{y})
\notin \{\phi(a):a\in\Sigma\}^\oplus$,
the only problem being that we never know whether it is still
worth continuing the approximation. So the situation would be similar
to a recursive enumeration (as opposed to a terminating decision
procedure). Unfortunately, this is \textbf{wrong}, and we want to underline
this fact to not give a misleading impression of the power of the methods
developed here:

\begin{thm}\label{limits}
There exists $\langle \Sigma,\phi\rangle\in\SA_{\SyS},
(\overline{x},\overline{y})$, such that
1. $(\overline{x},\overline{y})\notin\{\phi(a):a\in\Sigma\}^\oplus$, and yet
2. for every finite prefix $(x,y)=\sigma_n(\overline{x},\overline{y})$, 
$\{\phi(a):a\in\Sigma\}^\oplus \cap f_\omega(x,y)\neq \emptyset$.
\end{thm}

\proofbeg
Assume that the above claim is wrong, which means:
if $(\overline{x},\overline{y})\notin \{\phi(a):a\in\Sigma\}^\oplus$,
then there is a pair of finite strings
$(x,y)=\sigma_n(\overline{x},\overline{y})$
and $f_\omega(x,y)\cap \{\phi(a):a\in\Sigma\}^\oplus=\emptyset$.
We show how to construct
a decision procedure for checking the reachability problem
out of this fact, contradicting theorem \ref{undec}.

We have a recursive enumeration of 
$\{\phi(a):a\in\Sigma\}^\oplus$. So we can 
by turns enumerate an element of $\{\phi(a):a\in\Sigma\}^\oplus$, and
compute $\{w:\phi(w)\cap f_\omega(x,y)\neq\emptyset\}$ 
for a growing prefix $(x,y)$ of $(\overline{x},\overline{y})$. 
By this method, either at some point we will find
$(\overline{x},\overline{y})$ in our enumeration, 
or we will, for some finite $(x,y)\in(\Omega^*)^2$, find that
$\{w:\phi(w)\cap f_\omega(x,y)\neq\emptyset\}=\emptyset$,
meaning that $\{\phi(a):a\in\Sigma\}^\oplus\cap f_\omega(x,y)=\emptyset$. 
This entails that  $\{\phi(a):a\in\Sigma\}^\oplus$ is 
recursive -- contradiction.
\proofend

This is a negative result, but keep in mind 
we have never promised a decision procedure, just a method for
approximations! Now we can also state more precisely 
what it means for us to approximate a binary problem: 
if $(\overline{x},\overline{y})\notin\{\phi(a):a\in\Sigma\}^\oplus$, 
then the set of candidate strings will diminish for 
$\sigma_n(\overline{x},\overline{y})$ as $n$ increases; so we 
get closer to the empty set, though it is not said we reach it after
a finite number of steps (theorem \ref{undec}). Note that we can still be
more precise: the reachability problem is undecidable only if for all $n\in\mathbb{N}$,
$|\{w:\phi(w)\cap \sigma_n(\overline{x},\overline{y})\}|=\infty$
(otherwise we can easily check by brute force).

Now there is one substantial
concern which has to be addressed: could it  be that approximations
of the above kind are trivial, in the sense they can always be
performed in one way or another?
Assume we want to check whether 
$(\overline{x},\overline{y})\notin\{\phi(a):a\in\Sigma\}^\oplus$.
Then we just
take an enumeration of $\Sigma^*$, and check for each $w$
whether $(\overline{x},\overline{y})\in\phi(w)$. 
This way, we get a growing set of
strings we can exclude, and \textit{in the limit}, we find
a solution to our problem.
There is however a substantial difference between this
kind of approximation and the method we are proposing:
if $(\overline{x},\overline{y})\notin\{\phi(a):a\in\Sigma\}^\oplus$,
the latter enumeration method 
\textit{can never succeed}, because after any finite number of steps,
we will necessarily remain with an infinite set of strings to check.
So because this method comes with no hope
of success, it  is completely useless.
On the other side, in our case, we know that
approximation \textit{can} succeed: because each step of approximation
excludes a possibly infinite language of candidate strings. This does
of course not mean that it necessarily succeeds: then
it would be a decision procedure. This is, in other words, what
we mean by an approximation to the binary problem of reachability, and
this is the meaning of our results. 
We try to make this more precise in the following section.

\section{The Prospects of Approximation}

\subsection{Three Potential Applications of Theorem \ref{approx}}

Apart from this negative result, there are many upsides to our notion of
approximation. In particular, there are three different ways to
``interpret" theorem \ref{approx}, which are based on different interpretations of states, 
namely as atomic entities, as encoding objects 
in a space with a real-valued distance, and
as vectors of values of program variables.
One should keep in 
mind that sketching methods for $\SyS$,
they are \textit{a fortiori} applicable to transition relations which can be simulated
by $\SyS$ such as those of pushdown automata and many other methods in use.

The first interpretation of the notion of approximation is based on
states as discrete objects.
Given a semi-automaton $\langle \Sigma,\phi\rangle$,
a relation $R$, 
we put $\chi_\phi(R):=\{w\in\Sigma^{*}:\phi(w)\cap R\neq\emptyset\}$.
Given an SCA $\GA=\langle\Sigma,\phi,F_R\rangle\in\SA_{\SyS,\FIN}$, we can use the
methods described here to \textit{approximate} $L(\GA)$ (note that the restriction
that $F_R\in\FIN$ is not strictly necessary, as $\sigma_n[F_R]$ is finite anyway, 
but it simplifies things). We can obviously compute 
$f_\omega\circ \sigma_n[F_R]$. 
We then know that for any $\langle \Sigma,\phi,F_R\rangle\in\SA_{\SyS,\FIN}$,
$\chi_\phi(f_\omega\circ \sigma_n[F_R])$ is a regular language which is 
effectively computable (a finite union of regular languages).
Consequently, the sequence $(\chi_\phi(f_\omega\circ \sigma_n[F_R]))_{n\in\mathbb{N}}$ 
is a decreasing sequence of regular languages, each containing $L(\GA)$;
that is, $\chi_\phi(f_\omega\circ \sigma_n[F_R])\subseteq \chi_\phi(f_\omega\circ \sigma_{n+1}[F_R])	\subseteq L(\GA)$  
for all $n\in\mathbb{N}$.

\begin{defn}
Given an SCA $\GA=\langle\Sigma,\phi,F_R\rangle$,
 we define its $n$-\textbf{approximation}
$\GA_n=\langle\Sigma,\phi,f_\omega\circ \sigma_n[F_R]\rangle$.
\end{defn}

This kind of approximation is of course of little interest if we consider
the membership problem, which is decidable for $\SA_{\SyS,\FIN}$ anyways. However,
it might be very interesting as soon as we consider problems which
are undecidable for $\SA_{\SyS,\FIN}$. In the prior examples, we have
focussed on reachability; emptiness is related in the obvious way stated in lemma
\ref{reach}. But we can also consider problems like inclusion and universality.
As basically all important decision problems are decidable for the regular
languages, we can obviously compute these problems for arbitrary 
approximations $\GA_n$. So we get the following in a straightforward fashion:

\begin{lem}
Given $\GA,\GA'\in \SA_{\SyS,\FIN}$, we can decide whether
$L(\GA_n)=\emptyset$, $L(\GA_n)=\Sigma^*$, $L(\GA_n)\subseteq L(\GA'_n)$
for arbitrary $n\in\mathbb{N}$.
\end{lem}

This is clear because all are regular languages we can effectively compute, and
we approximate all these decision problems in this sense.
What we have to show is that these approximations are arbitrarily precise:
let $(L_n)_{n\in\mathbb{N}}$ be a decreasing sequence of languages,
that is $L_n\supseteq L_{n+1}$. We denote by $lim_{n\rightarrow\infty}(L_n)$
the unique language $L$ such that 1. for all $n\in\mathbb{N}$, $L_n\supseteq L$, and
2. if for all $n\in\mathbb{N}$, $L_n\supseteq L'$, then
$L'\subseteq L$.
This language always exists, and can be simply defined as 
$\{w:\forall n\in\mathbb{N}:w\in L_n\}=\bigcap_{n\in\mathbb{N}}L_n$.
The following lemma
establishes the desired correlation:

\begin{lem}
$lim_{n\rightarrow\infty}L(\GA_n)=L(\GA)$.
\end{lem}

\proofbeg
$\supseteq$ is obvious, as for all $n\in\mathbb{N}$, $L(\GA_n)\supseteq L(\GA)$,
and $lim_{n\rightarrow\infty}L(\GA_n)$ is the largest language satisfying
this condition.

$\subseteq$: assume $(\overline{x},\overline{y})\in f_\omega\circ \sigma_n[F_R]$ for all $n\in\mathbb{N}$.
After some finite number of positions (say $k$), all letters in $(\overline{x},\overline{y})$ 
are $\square$,
and for $(\overline{x}',\overline{y}')\in F_R$ the same holds (say after  $k'$ positions). Consequently, if the two agree
on the first $max\{k,k'\}$ positions, they will agree on all. Hence, we will have
$(\overline{x},\overline{y})\in F_R$, and if for all $n\in\mathbb{N}$,
 $\phi(w)\cap f_\omega\circ \sigma_n[F_R]\neq\emptyset$, then $\phi(w)\cap F_R\neq\emptyset$;
consequently, if $w\in L(\GA_n)$ for all $n\in\mathbb{N}$,
then $w\in L(\GA)$.
\proofend

Note however that this limit-construction
does not need to preserve \textit{any} computational properties:
we can easily construct a decreasing sequence or regular languages
$(L_n)_{n\in\mathbb{N}}$ such that $lim_{n\rightarrow\infty}L_n$
is not recursively enumerable: let $L\subseteq\Sigma^*$ be a language
which is not recursively enumerable, put 
$L_1=\Sigma^*$, and $L_{n+1}=L_n-\{w\}$ for some $w\in L_n$.
Then for all $n\in\mathbb{N}$, $L_n$ is a co-finite language and therefore
regular, whereas $lim_{n\rightarrow\infty}(L_n)=L$.

We now sketch how we might use this approximation to introduce  
a limit value which has some similarity to a probability, though the
values involved are clearly not probabilities in a technical sense. 
We let $\mathcal{I}$ denote the inclusion
problem, such that $\pr(\GA,\GA')=1$ if $L(\GA)\subseteq L(\GA')$,
and $\pr(\GA,\GA')=0$ otherwise. As we have said, this allows us to encode
the problem of emptiness and universality. We can now simply define a 
limit for stepwise approximation as follows:
\begin{center}
$Lim(\pr(\GA,\GA')=1)=lim_{n\rightarrow\infty}\frac{|\{i:i\leq n\&\pr(\GA_i,\GA'_i)=1\}|}{n}$ (provided this limit exists)
\end{center}
That is, the chances that $\pr(\GA,\GA')$ has a positive answer
would be defined as the limit of the cardinality of numbers $\leq n$ where it has a positive answer for $\GA_n,\GA'_n$, divided by $n$. 

Note that it is not said that our limit gives the correct answer to the
problem: take the case of $\GA_\emptyset$, which is an 
automaton $\langle\Sigma,\phi,\emptyset\rangle$ 
recognizing the empty language, and $\GA\in\SA_{\SyS,\FIN}$, which is such that
1. $L(\GA)=\emptyset$, and 2. for all $n\in\mathbb{N}$, $L(\GA_n)\neq\emptyset$
(this automaton must exist, otherwise emptiness would be decidable).
Now if we consider $\pr(\GA,\GA_\emptyset)$, we find that
for all $n\in\mathbb{N}$, $\pr(\GA_n,(\GA_\emptyset)_n)=0$, hence
$Lim(\pr(\GA,\GA_\emptyset))=0$, whereas $\pr(\GA,\GA_\emptyset)=1$.
So the  limit does not necessarily 
coincide with the correct answer to the problem!

However, in the case of universality, things work out better: 
let $\GA_{\Sigma^{*}}$ be a finite automaton such that
$L(\GA_{\Sigma^{*}})=\Sigma^{*}$, and hence for all $n\in\mathbb{N}$,
$L((\GA_{\Sigma^{*}})_n)=\Sigma^{*}$. Let $\GA$ be just any
automaton in $\SA_{\SyS,\FIN}$. It is easy to see that
in this case, we have $Lim(\pr(\GA_{\Sigma^{*}},\GA))=\pr(\GA_{\Sigma^{*}},\GA)$.
This is due to the fact that in
our approach, we always approximate from the top. 
We can slightly generalize this concept: 

\begin{defn}
An instance of the inclusion problem $\pr(\GA,\GA')$ is \textbf{correctly approximated},
if $Lim(\pr(\GA,\GA'))$ exists and $Lim(\pr(\GA,\GA'))=\pr(\GA,\GA')$.
\end{defn}

\begin{lem}
Let $\mathfrak{F}$ be a finite SCA.\footnote{By this, we mean an SCA such that
$\phi[\Sigma]$ is finite (mod $\eta$). This is obviously equivalent to a finite
state automaton.} Then the problem
$\pr(\mathfrak{F},\GA)$ is correctly approximated,
whereas $\pr(\GA,\mathfrak{F})$ in general is not.
\end{lem}

\proofbeg
The second part already follows from the counterexample above. 
For the first part, consider that for some $k\in\mathbb{N}$ and for all $n>k$ 
we have $L(\mathfrak{F}_n)=L(\mathfrak{F})$.  
So if $\pr(\mathfrak{F},\GA)=1$, then for all $n>k$, $\pr(\mathfrak{F}_n,\GA_n)=1$, 
and so  $Lim(\pr(\mathfrak{F},\GA))=1$.

Conversely, assume  $Lim(\pr(\mathfrak{F},\GA))=1$. Then if $w\in L(\mathfrak{F})$, then
$w\in L(\GA_n)$ for all $n\in\mathbb{N}$. As $L(\GA)=lim_{n\rightarrow\infty}L(\GA_n)$,
it follows that $w\in L(\GA)$, so $\pr(\mathfrak{F},\GA)=1$.
\proofend

Note that by this result, we have \textit{a fortiori} the negative result that
$\pr(\GA,\GA')$ cannot be correctly approximated in the general case where
$\GA,\GA'\in \SA_{\SyS,\FIN}$. Moreover, in this case it is not clear whether the limit even
exists. From the fact that the universality problem can be correctly approximated,
it follows that even if the limit exists, in general it is not computable.
We can, however, compute 
$Lim_n(\pr(\GA,\GA')=1)=\frac{|\{i:i\leq n\&\pr(\GA_i,\GA'_i)=1\}|}{n}$ for any $n$.
This way, we might be able to reasonably estimate the chances that
a certain instance of the inclusion problem has a certain answer, at 
least if the problem is correctly approximated. 

To make clear what about these results is peculiar to $\SyS$, we
should add the following: we can provide
limit constructions for approximations of reachability for any recursive
class of relations, for example by taking the limit of reachability with words of length $\leq n$. The difference of this approach to the ones we sketched here
is that the former are always confined to finite stringsets; we never make the
step to \textit{infinite} languages approximating the target language. This
is a fundamental shortcoming, as we can never talk about infinitary properties. On the other side, the two methods might be combined: as we have
seen, our approximation of languages is always ``from the top", proceeding
to smaller languages, whereas a concept as ``reachability with words of 
length $\leq n$" always comes from the bottom. 


The following two
interpretations of theorem \ref{approx} are more genuine to $\SyS$, and
there is no way to get similar results without the fundamental properties of
synchronous subsequential relations.
We  first provide an interpretation of states which allows for
a numeric approximation. It is based on the concept
of state-strings being allocated in a space, where for each two strings
we have a unique real-valued distance.
A distance function on $\Omega^*$ is any function
$f:\Omega^*\times\Omega^*\rightarrow \mathbb{R}^+_0$, such that
$f(w,w)=0$ for all $w\in\Omega^*$, and $f(w,v)=f(v,w)$.
A distance function measures \textit{how close} a string $w$ is
to another string $v$.

This is a very general notion; 
we will consider more restrictive distance functions
for the set $\{\eta(\overline{w}):\overline{w}\in\Omega^\omega\}$. 
By $\gcp(w,v)$ we denote the \textit{greatest common prefix} of $w,v$, that is,
$\gcp(aw,av)=a(\gcp(w,v))$, and $\gcp(aw,bv)=\epsilon$ if $a\neq b$.
A \textbf{normal distance function} on $\Omega^*$ is defined as follows: 
\begin{enumerate}
\item there is a map $val:\Omega^{+}\rightarrow \mathbb{R}$, such that
if $\square^{n}\in \pref(w)$, $\square^{n}\notin \pref(v)$, then
$val(w)>val(v)$.
\item For  strings $w,w',v$, if $|\gcp(w,v)|> |\gcp(w',v)|$, then
$dist(w,v)< dist(w',v)$ -- that is, the longer the common prefix,
the smaller the distance.
\item If $v=wx$, then $dist(w,v)=val(x)$.
\end{enumerate}
Note that 3. implies that $val(\epsilon)=0$ and $dist(\epsilon,w)=val(w)$.
Therefore condition 1. only applies to $\Omega^+$.
An easy example of such a measure is if we let words represent real numbers
in $[0,1]$ (in $|\Omega|$-ary representation). We simply  specify a
linear order $<$ on $\Omega$. Then $val$ is just 
the map from the representation to the number
it represents, and $dist(x,y)$ is defined by $|val(x)-val(y)|$. Note that our definition
is independent on the order $<$ in $\Omega$!
Given a normal distance measure
$dist$, we call a \textbf{convex region} of $\Omega^*$ a set of words
$M$, such that $M=\{w:dist(v,w)\leq x$ for some $v\in\Omega^*$, $x\in\mathbb{R}\}$;
we also say $M$ has \textit{center} $v$. The following properties
are easy to see:
\begin{enumerate}
\item For every normal distance function $dist$,
$n\in \mathbb{N}$, $\overline{w}\in \Omega^\omega$, $ \eta[ f_\omega\circ \sigma_n(\overline{w})]$
forms a convex region in the space defined by $dist$. 
We call such a space a \textbf{normal
subspace} of $\Omega^*$.
\item Given two normal subspaces $X,Y\subseteq\Omega^{*}$ and
$\langle\Sigma,\phi\rangle\in\SA_\SyS$, $\chi_\phi(X\times Y)$ is a
regular language. This follows from theorem \ref{approx}, which is basically
a statement on normal subspaces.
\item For any two $x,y\in\Omega^*$, $\langle\Sigma,\phi\rangle\in\SA_\SyS$, 
we can effectively compute the 
language $\chi_\phi(X\times Y)$ for arbitrarily
small normal subspaces $X,Y$ with center $x$ and $y$, respectively;
moreover, this language is regular.
\end{enumerate}

So there is an interpretation where our notion of approximation is
very useful: if we consider a string as representing a unique numeric value,
we \textit{cannot} determine whether the system can reach value
$y$ from value $x$; but we can determine, for any $\epsilon>0$,
whether we can reach the interval $[y-\epsilon,y+\epsilon]$ from the
interval $[x-\epsilon,x+\epsilon]$. But of course, approximation
works in scenarios which are much more general than the interpretation
of strings as real numbers. What is particularly interesting about this
approach is that it provides a numeric approximation in a purely
symbolic setting. 

There is a third interpretation which is based on the encoding of
program semantics as strings. As we have said, in order to be able to model a program
in terms of relations in $\SyS$, we need a  linear hierarchy of 
variables which can be thought of in terms of importance: because influence 
goes only in one direction, the more to the left a variable is encoded in the
string, the more important it is. 
Now the approximation
means: we reach  the desired configuration at least as regards the $n$-\textit{most
important parameters}, for $n\rightarrow\infty$. Moreover, we can
effectively compute the set of sequences of computation steps which yield this result.

\subsection{Complexity Issues}

There is one most fundamental problem regarding the complexity of our approximation 
techniques, which is the following: given a finite set $\textbf{R}\subseteq \SyS$ of
relations, what is the complexity for computing $\sigma_n[\textbf{R}^\oplus]$?
On the positive side, we know from lemma \ref{comm} that
$\sigma_n[\textbf{R}]^\oplus=(\sigma_n[\textbf{R}])^\oplus$;
so the problem is surely computable, as $\sigma_n[\textbf{R}]$ is a finite set of finite
relations. $\sigma_n[\textbf{R}]$ is rather easily computed in some way or other;
the difficult thing is to compute the composition-closure. As $\sigma_n(R)$ is always
a finite relation, this will be computable in a finite number of steps. The problem
is to find the smallest $n$ such that $\bigcup_{m\leq n}\{R_i:i\in I\}^{m}=\{R_i:i\in I\}^{\oplus}$. The bad thing is that this $n$ depends on the size of 
$M$, where $M$ is the underlying set from which tuple components in the relation
are taken. For example, consider the relation 
$S_k=\{(n,n+1):n\in \mathbb{N},n< k\}$ for some $k\in\mathbb{N}$.
This is a finite relation; still to get $\{S_k\}^{\oplus}$, we need no less than $k$
iterated compositions. In our case, the underlying set is $\Omega^{n}$ for some alphabet
$\Omega$, 
and it is easy to see that this grows exponentially by factor $|\Omega|$
in terms of $n$. So assume for all $R\in \textbf{R}$, we have $R\subseteq
(\Omega\times\Omega)^{*}$; then in the worst case, in order to compute
$\sigma_n[\textbf{R}]^\oplus$ we need at least $|\Omega|^{n}$ computation
steps. 
Note that this is only the most fundamental of all problems, which
is necessary in order to compute for example $\chi_\phi(f_\omega\circ\sigma_n[F_R])$
etc. This is where the notions of direct and subdirect product become interesting,
as for relations relations being (sub)directly decomposable, the approximation is
much easier. This is however beyond our current scope.

The fact that computing approximations is exponential in terms of $n$
is discouraging, so our preliminary results may not be fully 
satisfying. Still, in practical application many problems are not as hard
as in theory (which always takes the worst case). In the theory
of infinite automata, there is a necessary trade-off between
expressiveness on the one hand and decidability issues on the
other. Our notion of $\SyS$ relations and results on approximation,
other than exploring the space of possibilities, might serve
to establish some reasonable position in between the two.

%
%
%

\section{Conclusion}

In this paper, we have introduced the class $\SyS$ of synchronous 
subsequential relations and investigated its properties for
the theory of infinite automata.  $\SyS$ 
covers many possible models of computation, such as simple or embedded stacks
(as in \cite{weir:control}) and many more. The underlying intuition is that
there is a linear hierarchy of program variables such that computations performed on higher variables affect computations on lower variables, but not
vice versa.
Though automata with primitive
transitions relations in $\SyS$ have an undecidable reachability (and emptiness,
universality and inclusion)
problem, there is a way to approximate the problem which
does not seem to work for more expressive classes such as the regular
or subsequential relations, as it presupposes both a 1-1 correspondence of input- and
output-letters, and independence of computations of later computation steps.
The method in itself seems to be of some interest and is not necessarily
bound to $\SyS$ (though it requires some fundamental properties of $\SyS$), 
so one might further investigate on it. 
We have given a sketch of how one might compute the chances
that an instance of a decision problem has a certain answer, and how the
problem can be approximated in the sense of numeric distances and the
hierarchy of program variables. 
Finally, the methods presented here are based 
on (algebraic) decompositions of relation
monoids. There seems to be rather little work in this vein, so we hope
the application of (relation-)algebraic methods to the theory of
infinite automata, to which
we have laid some fundamentals in this paper, might open the road
to further interesting results.

\bibliographystyle{eptcs}
\bibliography{alles_7_14}

\end{document}